\documentclass[11pt]{article}

\usepackage{latexsym,color,graphicx,amsmath,amssymb}
\usepackage{url}

\newtheorem{theorem}{Theorem}[section]

\def\reals{{\mathbb R}}
\def\cplx{{\mathbb C}}
\def\xx{{\bf x}}

\def\vv{{\bf v}}
\def\xx{{\bf x}}

\begin{document}

\title{Homotheties and incidences\thanks{%
  Work by Micha Sharir has also been supported by
  Grant 338/09 from the Israel Science Fund,
  by the Israeli Centers of Research Excellence (I-CORE) program
  (Center No.~4/11), by the Blavatnik Research Fund in Computer Science at Tel Aviv University,
  and by the Hermann Minkowski--MINERVA Center for Geometry at Tel Aviv
  University.
}}

\author{
Dror Aiger\thanks{%
  Google Inc.;
  email: {\tt aigerd@google.com}}
\and
Micha Sharir\thanks{%
  School of Computer Science, Tel Aviv University, Tel~Aviv 69978, Israel;
  email: {\tt michas@tau.ac.il}}
}


\maketitle

\begin{abstract}
We consider problems involving rich homotheties in a set $S$ of $n$ points in the plane
(that is, homotheties that map many points of $S$ to other points of $S$).
By reducing these problems to incidence problems involving points and lines in $\reals^3$,
we are able to obtain refined and new bounds for the number of rich homotheties, and for 
the number of distinct equivalence classes, under homotheties, of $k$-element subsets of 
$S$, for any $k\ge 3$. We also discuss the extensions of these problems to three and higher dimensions.
\end{abstract}

\section{Introduction}

In this note we extend the analysis technique of Guth and Katz~\cite{GK2}, which is based on
the framework proposed by Elekes (as exposed, e.g., in \cite{EKS}), 
to handle homotheties of point sets in the plane.
The original technique in \cite{GK2} derives a lower bound on the number of distinct distances
determined by a set $S$ of $n$ points in $\reals^2$. Equivalently, this can be thought of as 
obtaining a lower bound on the number of equivalence classes of pairs in $S\times S$ under 
Euclidean motions. That is, two pairs $(a,b)$, $(a',b')$ are equivalent if there exists a 
rigid motion that maps $a$ to $a'$ and $b$ to $b'$ (which is the same as saying that $|ab|=|a'b'|$).

In this note we consider the analogous problems that arise when we replace rigid motions by
homotheties. Both rigid motions and homotheties have three degrees of freedom, and so can be
represented as points in parametric 3-space. As we note here, Elekes's transformation
 can be applied in the context of homotheties too, and 
reduce problems that involve homotheties acting on a finite point set in the plane to problems that 
involve incidences between points and lines in three dimensions. The machinery developed in \cite{GK2} 
then allows us to obtain various results concerning homotheties in the plane.

\paragraph{Our results.}
Specifically, we show that the number of \emph{$t$-rich homotheties} in a set $S$ of $n$ points in
the plane, namely, homotheties that map at least $t$ points of $S$ to other points of $S$, is 
${\displaystyle O\left( \frac{n^3}{t^2} + \frac{n^2\nu^2}{t^3} \right)}$, where $\nu$ is the maximum size of 
a collinear subset of $S$. We also show, via a simple construction, that the second term cannot be improved. 
The upper bound is a consequence of a general incidence bound (given in Theorem~\ref{hominc} below) 
between any set of $m$ homotheties (represented as points in $\reals^3$) and $N$ lines of the form 
$h_{p,q}$, for pairs $p,q\in S$, where $h_{p,q}$ is the locus of all homotheties that map $p$ to $q$ 
(which is indeed a line under a suitable and natural parameterization of homotheties---see below).

We then use this bound to obtain a lower bound on the number of pairwise non-homothetic $k$-tuples in
$S$. The bound, given in Theorem~\ref{dist:hom}, is $\Omega(n^{k-1})$ for $k\ge 4$.
For $k=3$ the bound depends on the parameter $\nu$: it is $\Omega(n^2)$ if $\nu = O(n/\sqrt{\log n})$, 
and is $\Omega\left( \frac{n^4}{\nu^2\log n} \right)$ otherwise.
As noted in \cite{PS}, the bound is worst-case tight for $k\ge 4$; it improves an earlier bound
of $\Omega(n^{k-2})$. For $k=3$ an easy upper bound is $O(n^2)$, so our bound is optimal in this 
case too when $\nu = O(n/\sqrt{\log n})$.

\paragraph{Background.}
Problems involving distinct equivalence classes among $k$-tuples of point sets, under various transformations,
have been posed and investigated quite some time ago. They are mentioned, e.g., in the monographs
Brass et al.~\cite{BMP} and Pach and Sharir~\cite{PS}. Many types of transformations have been considered,
including rigid motions, homotheties, similarities, all the way to general affine and projective transformations.
The simplest case is where $k=2$ and the transformations are rigid motions. In this case,
as already noted, the question is how many
distinct distances are determined by any set of $n$ points in the plane. This problem has been almost completely
settled in Guth and Katz~\cite{GK2}, who derived the lower bound $\Omega(n/\log n)$ for this quantity, almost
matching Erd{\H o}s's upper bound $O(n/\sqrt{\log n})$.

Guth and Katz's solution is based on an ingenious tranformation due to Elekes (exposed, e.g., in \cite{EKS}),
which considers a standard representation of rigid motions in the plane as points in parametric 3-space,
and maps each pair $p,q$ of points in the input set $S$ to a line $h_{p,q}$, which is the locus of all rigid 
motions that map $p$ to $q$. (With the right choice of parameterization, given in \cite{GK2}, $h_{p,q}$ is 
indeed a line.) By deriving new and sharper upper bounds on the number of incidences between points and lines 
in $\reals^3$, Guth and Katz were able to obtain upper bounds on the number of \emph{$t$-rich rigid motions}, namely motions
that map at least $t$ points of $S$ to other points of $S$. These bounds, combined with another reduction 
of Elekes, yield the aforementioned lower bound on the number of distinct distances.

Rudnev~\cite{Rud} has later noticed that the same general approach can also handle the case $k=3$
for rigid motions, i.e., yield a lower bound on the number of pairwise non-congruent triangles determined by $S$,
and also the case of $k=3$ under similarities. Rudnev obtained the lower bound $\Omega(n^2)$ in the former
case, and the lower bound $\Omega(n^2/\log n)$ for the number of pairwise non-similar triangles.

However, as Rudenv was also aware of, the case of similarities has already been handled by Solymosi and Tardos~\cite{SoTa}, before
the new algebraic machinery of Guth and Katz came into play. We will also comment on how to handle similarities,
in the interest of completeness, but, similar to Rudnev, we do not see a way in which the algebraic technique,
powerful as it is, can improve the bound in \cite{SoTa}.

We therefore focus in this note on homotheties, which seem to have received less attention, and apply the 
algebraic machinery to obtain the aforementioned results. Problems involving
homotheties have been posed in several earlier works, including Brass~\cite{Br}, Brass et al.~\cite{BMP},
and Pach and Sharir~\cite{PS}. The earlier works (going back to van Kreveld and de Berg~\cite{vKdB}) 
have mostly considered the ``complementary'' question
of bounding the maximum number of $k$-subsets of a set $S$ of $n$ points in the plane that are homothetic to
a given $k$-element ``pattern'' $P$. Elekes and Erd{\H o}s~\cite{EE} (see also Brass~\cite{Br})
have shown that this quantity is $\Theta(n^{1+1/k})$, where $k$ is the dimension of the rational 
affine hull of $P$. This becomes $\Theta(n^2)$ for (only) one-dimensional patterns, and can be 
attained only under certain algebraicity assumptions, as shown in Laczkovich and Ruzsa~\cite{LR}.
The algorithmic issues of finding the homothetic copies of $P$ in $S$ are discussed in \cite{Br,vKdB}.

Various open problems involving homotheties in three and higher dimensions are mentioned in \cite{BMP,PS}.

\section{Homotheties in the plane}

Each homothetic transformation of the plane (translation and scaling) has three degrees of freedom,
and can therefore be represented parametrically as a point in $\reals^3$. 
Let us use the representation $(\xi,\eta,t)$, where the homothety first scales the plane by $t$, 
with respect to the origin, and then translates it by the vector $(\xi,\eta)$.
That is, the point $(\xi,\eta,t)$ represents the transformation $\tau_{\xi,\eta,t}(x) = tx + (\xi,\eta)$,
for $x\in\reals^2$.

For a pair of points $p,q\in\reals^2$, the locus in $\reals^3$ of all homotheties $(\xi,\eta,t)$
that map $p=(p_x,p_y)$ to $q=(q_x,q_y)$ is the line $h_{p,q}$, given by the equations
\begin{align*}
\xi & = q_x -tp_x \\
\eta & = q_y -tp_y .
\end{align*}
Let $S$ be a set of $n$ points in the plane, and let $L$ denote the collection of all 
$n^2$ lines $h_{p,q}$, for $p,q\in S$.

A homothety $(\xi,\eta,t)$ is incident to $t$ lines $h_{p_i,q_i}$ of $L$, for $i=1,\ldots,t$, 
if and only if it maps $p_i$ to $q_i$ for each $i$. In other words, we can reduce questions
about homotheties acting on $S$ to questions about incidences between points and lines in three dimensions.

The latter problem has been studied in Guth and Katz~\cite{GK2}. The bound that they obtain,\footnote{%
  Technically, this bound is implicit in, but directly follows from their results.}
for the number of incidences between $M$ points and $N$ lines in $\reals^3$ is
\begin{equation} \label{gk:inc}
I(M,N) = O\left( M^{1/2}N^{3/4} + M^{2/3}N^{1/3}s^{1/3} + M + N \right) ,
\end{equation}
where $s$ is the maximum number of input lines that lie in a common plane.

To apply (\ref{gk:inc}) in our context, we estimate the parameter $s$, in our scenario, as follows.
A plane $\pi$ in $\reals^3$, with equation $\xx\cdot\vv = c$, for a vector 
$\vv=(v_1,v_2,v_3)$ and a real $c$, contains the line $h_{p,q}$ if
$$
(q-tp)\cdot (v_1,v_2) + tv_3 = c 
$$
holds for every $t$. This happens if and only if $p$ and $q$ lie on the two respective parallel lines (in $\reals^2$)
\begin{align*}
p\cdot (v_1,v_2) & = v_3 \\
q\cdot (v_1,v_2) & = c .
\end{align*}
That is, in order to apply the incidence bound of Guth and Katz~\cite{GK2}, we have to control
configurations with many points lying on pairs of parallel lines. Actually, since the two lines do not have
to be distinct, the parameter that we are after is the maximum number of points of $S$ on any single line.
We denote this quantity as $\nu(S)$, and put
$$
\mu(S) := \max \left\{ |S\cap\ell|\cdot |S\cap\ell'| \mid \text{$\ell$ and $\ell'$ parallel lines} \right\} = \nu^2(S) .
$$
The preceding reasoning then implies that no plane in $\reals^3$ contains more than $\mu=\mu(S)$ 
lines of $L$. Applying the bound in (\ref{gk:inc}), we then get the following result.
\begin{theorem} \label{hominc}
Let $S$ be a set of $n$ points in the plane, and put $\nu=\nu(S)$. Let $L$ be the set of
the $n^2$ lines $h_{p,q}$, for $p,q\in S$, in $\reals^3$, and let $H$ be a set of $m$ homotheties
of the plane, represented as points in $\reals^3$. Then the number of incidences between the
points in $H$ and the lines in $L$ satisfies
\begin{equation} \label{hom:inc}
I(H,L) = O\left( m^{1/2}n^{3/2} + m^{2/3}n^{2/3}\nu^{2/3} + m + n^2 \right) .
\end{equation}
An analogous bound holds if we replace $L$ by any subset $L'$; the bound is then
$$
I(H,L') = O\left( m^{1/2}|L'|^{3/4} + m^{2/3}|L'|^{1/3}\nu^{2/3} + m + |L'| \right) .
$$
\end{theorem}
In particular, the number $M_{\ge t}$ of \emph{$t$-rich} homotheties, namely those
that map at least $t$ points of $S$ to other points of $S$, satisfies
\begin{equation} \label{gek}
M_{\ge t} = O\left( \frac{n^3}{t^2} + \frac{n^2\nu^2}{t^3} + \frac{n^2}{t} \right) =
O\left( \frac{n^3}{t^2} + \frac{n^2\nu^2}{t^3} \right) ,
\end{equation}
since the first term always dominates the third one. This bound follows in a standard manner
by denoting by $H_{\ge t}$ the set of $t$-rich homotheties, of cardinality $m=M_{\ge t}$,
and by combining the bound in (\ref{hom:inc}) with the inequality $I(H_{\ge t},L) \ge tM_{\ge t}$. 
Of the remaining two terms, the first (resp., second) term dominates when $\nu^2 \le nt$ (resp., $\nu^2 \ge nt$). 

\paragraph{Lower bound.}
We next show that the bound in (\ref{gek}) is tight in the worst case when $\nu \ge \sqrt{nt}$,
that is, when the second term in (\ref{gek}) dominates.
Assuming\footnote{%
  For $n<t$ there are no $t$-rich homotheties. For $t<n<16t$ the upper bound in (\ref{gek}) is $O(t)$,
  and a matching lower bound, using a set of $n$ equally spaced points on a line, is easy to derive
  (unless $t=n-o(n)$).} 
that $n\ge 16t$, we also have $\nu \ge 4t$. Construct the set
$$
S_0 = \{ (i,0) \mid i=1,\ldots,\nu \} ,
$$
put $t = n/\nu$, create $t$ translated copies of $S_0$, denoted as $S_1,\ldots,S_t$,
and let $S$ be the union of these translated copies.
We choose the translation vectors generically, to ensure that no non-horizontal line contains more
than two points of $S$, and that, for any homothety of the plane that maps two horizontal lines that contain 
copies of $S_0$ to two other such lines, one of the two source copies is such that none of its points are
mapped to points of the corresponding target copy. 
Clearly, $\nu(S) = \nu$. To obtain a homothety that maps at least $t$
points of $S$ to other points of $S$, we choose a copy $S_i$ of $S_0$, and choose an arithmetic
progression in $S_i$ of at least $t$ elements. To do so, choose the difference of the sequence 
to be any integer $1\le j < \nu/(2t)$, and start the sequence at the $i$-th element, for any $i<j$.
Denote the resulting sequence as $A$. Now pick another copy $S_{i'}$ of $S_0$, and choose in it
any pair of elements so that they are the first two elements of an arithmetic sequence $B\subseteq S_{i'}$ 
of length at least $t$, and its
difference is relatively prime to $j$. There are $\Omega(\nu\cdot(\nu/t)) = \Omega(\nu^2/t)$ such pairs
in $S_{i'}$, a bound that follows from standard properties of Euler's totient function 
(see, e.g., \cite[Lemma 6.17]{Ed} and \cite{HW}).
We now map $A$ to $B$ by a homothety.

We claim that all these homotheties are distinct. Indeed, each such homothety is uniquely determined
by the choice of the first two elements of $A$ and the first two elements of $B$. Now, for two homotheties
$\tau_1$, $\tau_2$ to coincide, they must use the same source copy $S_i$ and the same target copy $S_{i'}$
of $S_0$. Assume that $\tau_1$ (resp., $\tau_2$) is determined, as above, by $p,q\in S_i$ and $p',q'\in S_{i'}$ 
(resp., by $r,s\in S_i$ and $r',s'\in S_{i'}$);
to simplify the reasoning, we use these symbols to refer also to the $x$-coordinates of the
preimages of these points on $S_0$, with respect to the corresponding translations.
Putting $j=q-p$, $j'=q'-p'$, $a=s-r$, and $a'=s'-r'$, we thus have $\lambda = j'/j = a'/a$, where $\lambda$
is the scaling factor of the homothety $\tau_1=\tau_2$. Since $j,j'$ and $a,a'$ are both relatively prime,
it follows that $a=j$ and $a'=j'$. We now claim that $r=p$ too. If not, we have, by construction,
$p,r < j$. Since $p$ is mapped to $p'$ and $r$ to $r'$, we also have $\lambda = \frac{r'-p'}{r-p}$,
which contradicts the fact that $j$ and $j'$ are relatively prime (as $|r'-p'| < j'$ and $|r-p| < j$). 
It now follows that $(p,q,p',q') = (r,s,r',s')$, and we thus conclude that distinct quadruples of 
this kind determine distinct homotheties, as claimed.

The number of such homotheties is thus at least
$$
\Theta(t^2) \cdot \Theta\left( \left(\frac{\nu}{t}\right)^2 \cdot \frac{\nu^2}{t} \right) =
\Theta\left( \frac{n^2}{\nu^2} \cdot \frac{\nu^4}{t^3} \right) = 
\Theta\left( \frac{n^2\nu^2}{t^3} \right) .
$$
Unfortunately, we still do not know whether the bound (\ref{gek}) is tight also for the case $\nu < \sqrt{nt}$.

\paragraph{Lower bound for distinct homothety classes.}
Let $S$ be a set of $n$ points in the plane, and let $k\ge 3$ be an integer parameter.
Two ordered $k$-tuples $(a_1,\ldots,a_k)$, $(b_1,\ldots,b_k)$ are said to be 
\emph{equivalent} under a homothety if there exists a homothety that maps $a_i$ to $b_i$, for each $i$.
We want to obtain a lower bound on the number of distinct equivalence classes of $k$-element subsets
of $S$ under homotheties. This is done using the following variant of Elekes's tranformation.

Let $Q$ denote the set of all $2k$-tuples $(a_1,\ldots,a_k,b_1,\ldots,b_k)$ of elements of $S$,
with the $a_i$'s all distinct and the $b_i$'s all distinct, such that
$(a_1,\ldots,a_k)$ is equivalent to $(b_1,\ldots,b_k)$ under a homothety.
Let $x$ denote the number of distinct equivalence classes of $k$-tuples, 
and let $E_1,\ldots,E_x$ denote the classes themselves. Clearly, we have, by the Cauchy-Schwarz inequality,
$$
|Q| = \sum_{i=1}^x \binom{|E_i|}{2} =
\frac12 \sum_{i=1}^x |E_i|^2 - \frac12 \sum_{i=1}^x |E_i| 
\ge \frac{1}{2x} \left( \sum_{i=1}^x |E_i| \right)^2 - \frac12 \sum_{i=1}^x |E_i| 
= \Omega\left( \frac{n^{2k}}{x} \right) ,
$$
where the last inequality holds if we assume that $x\ll n^k$ (otherwise we get a better lower bound 
than the one we aim for---see the introduction and Theorem~\ref{dist:hom} below).

For an upper bound on $|Q|$, we note that every homothety that maps exactly $t$ points of $S$
to $t$ other points generates 
$$
t(t-1)\cdots (t-k+1) \le t^k
$$
elements of $Q$, and we thus have
$$
|Q| \le \sum_{t\ge k} t^k M_t = O\left( k^k M_{\ge k} + \sum_{t \ge k+1} t^{k-1} M_{\ge t} \right) ,
$$
where $M_t$ (resp., $M_{\ge t}$) is the number of homotheties that map exactly $t$ 
(resp., at least $t$) points of $S$ to other points of $S$.

Using the upper bound (\ref{gek}) on $M_{\ge t}$, we have
\begin{align*}
|Q| & = O\left( k^k \left( \frac{n^3}{k^2} + \frac{n^2\nu^2}{k^3} \right) 
+ \sum_{t \ge k+1} t^{k-1} \left( \frac{n^3}{t^2} + \frac{n^2\nu^2}{t^3} \right) \right) \\
& = O\left( n^3k^{k-2} + n^2\nu^2k^{k-3} 
+ \sum_{t \ge k+1} \left( n^3t^{k-3} + n^2\nu^2t^{k-4} \right) \right) .
\end{align*}
For $k=3$ the sum is $O(n^4 + n^2\nu^2\log n)$. Combining this with the lower bound 
$|Q| = \Omega(n^6/x)$, we obtain
$$
x = \begin{cases}
\Omega(n^2) & \text{if $\nu = O(n/\sqrt{\log n})$} \\
\Omega\left( \frac{n^4}{\nu^2\log n} \right) & \text{otherwise} .
\end{cases}
$$
The situation is simpler for larger values of $k$, in which case we have
$$
|Q| = O\left( n^{k+1} + n^{k-1}\nu^2 \right) = O\left( n^{k+1} \right) ,
$$
implying that $x = \Omega(n^{k-1})$. That is, we have:
\begin{theorem} \label{dist:hom}
The number of distinct equivalence classes of $k$-element subsets of a set $S$ of $n$ 
points in the plane, under homotheties, is $\Omega(n^{k-1})$ for $k\ge 4$.
For $k=3$ the lower bound depends on the maximum number $\nu$ of points of $S$
in any common line. It is $\Omega(n^2)$ if $\nu = O(n/\sqrt{\log n})$, and is
$\Omega\left( \frac{n^4}{\nu^2\log n} \right)$ otherwise.
\end{theorem}
The theorem solves Problem 6.1~in Pach and Sharir~\cite{PS}. 
As already noted, the bound is worst-case tight for $k\ge 4$, using a simple construction given in~\cite{PS};
the previously best known lower bound was $\Omega(n^{k-2})$ (see~\cite{PS} for the easy argument).
For $k=3$, the bound is worst-case tight when $\nu = O(n/\sqrt{\log n})$. We leave it as an open
problem to tighten the small remaining gap when $\nu$ is larger.

\paragraph{Joints.}
A \emph{joint} in $L$ is a point (homothety) that is incident to at least three non-coplanar lines of $L$.
By the preceding reasoning, if $\tau$ is a homothety incident to the non-coplanar lines
$h_{p_i,q_i}$, $i=1,2,3$, then $p_1,p_2,p_3$ is a non-collinear triple, and so is the triple
$q_1,q_2,q_3$, consisting of the respective images of $p_1,p_2,p_3$ under $\tau$.
The number of joints in $L$ is $O(|L|^{3/2}) = O(n^3)$~\cite{GK,KSS,Qu}.
That is, there are at most $O(n^3)$ homotheties that map at least three non-collinear points of $S$ 
to other (non-collinear) points of $S$.

We mention this result because it does not depend on the parameter $\nu(S)$. Note that the bound in
(\ref{gek}) is $O(n^3)$ only for $\nu(S) = O(n^{1/2})$. Another way to interpret this finding is that
in order to get many $t$-rich homotheties we need $S$ to contain many collinear points. Moreover, in
the lower bound construction given above, all the homotheties that we construct map (at least $t$)
collinear points of $S$ to other collinear points, and no other point of $S$ is mapped to a point of $S$.
The observation concerning joints, as just given, indicates that this is indeed unavoidable---the number 
of homotheties that map at least one non-collinear triple in $S$ to another such triple is much smaller.

\subsection{Homotheties in higher dimensions}

Unlike the case of distinct distances (that is, of equivalence classes under rigid motions),
Elekes's tranformation extends easily to higher dimensions in the case of homotheties. 
In $\reals^d$, a homothety has $d+1$ degrees of freedom, and can be represented by $(\xi,t)$,
where $t\in\reals^+$ is the scaling factor and $\xi\in\reals^d$ is the translation vector.
The locus $h_{p,q}$ of homotheties (as points in $\reals^{d+1}$) that map a point $p$ to 
another point $q$ in $\reals^d$ is still a line, given by the system $q = tp + \xi$ of $d$ 
linear equations in $d+1$ variables. Hence, the basic question that we face, analogous to 
the one studied in Theorem~\ref{hominc}, is to estimate the number of incidences
between points and lines in $\reals^{d+1}$.

While much simpler than the corresponding transformation for rigid motions, this is far
from being an easy problem, and it gets harder as $d$ increases. So far the only known
nearly tight bound for points and lines is in four dimensions (that is, for $d=3$), due to
Sharir and Solomon~\cite{SS4}. For a set $H$ of $m$ points (homotheties in our case) and 
a set $L$ of $N$ lines (the $n^2$ lines $h_{p,q}$, for $p,q\in S$, in our case), one has
\begin{equation} \label{plinc4}
I(H,L) = O\left( 2^{c\sqrt{\log m}} (m^{2/5}N^{4/5} + m) + m^{1/2}N^{1/2}q^{1/4} + m^{2/3}N^{1/3}s^{1/3} + N \right) ,
\end{equation}
for a suitable absolute constant $c$, provided that no 2-plane contains more than $s$ lines
of $L$ and that no hyperplane or quadric contains more than $q$ lines of $L$.

We will shortly use this bound to obtain an upper bound on the number of $t$-rich homotheties
in a set $S$ of $n$ points in three dimensions, which is better than the one in (\ref{gek})
for $t\ll n$ and for certain ranges of other parameters, discussed below. 
However, as it turns out, and perhaps surprisingly, this improved bound does not 
lead to an improved bound on the number of equivalence classes, under homotheties, 
of $k$-element subsets of $S$, for any $k\ge 3$. As we will show, the planar bounds, given
in Theorem~\ref{dist:hom}, are large enough, so that the improvement in the bound 
on the number of $t$-rich homotheties, an improvement that holds only when $t\ll n$,
does not lead to a similar improvement in the number of equivalence classes.

Nevertheless, for the sake of its own interest, we proceed to bound the number of $t$-rich homotheties.

\paragraph{Lines contained in planes, hyperplanes, or quadrics.}
In order to apply the bound in (\ref{plinc4}), we first proceed to understand the geometric structure 
of the parameters $q$ and $s$ in (\ref{plinc4}).

For estimating $s$, we note that a 2-plane $\pi$ is the intersection of two hyperplanes 
in $\reals^4$, given by equations of the form
\begin{align*}
(\xi,t)\cdot (v_1,u_1) & = c_1 \\
(\xi,t)\cdot (v_2,u_2) & = c_2 ,
\end{align*}
for suitable vectors $v_1$, $v_2\in\reals^3$ and scalars $u_1$, $u_2$, $c_1$, $c_2$. For a line $h_{p,q}$
to be contained in $\pi$, we must have
\begin{align*}
(q-tp)\cdot v_1 + tu_1 & = c_1 \\
(q-tp)\cdot v_2 + tu_2 & = c_2 ,
\end{align*}
for every $t$, meaning that $p$ and $q$ must lie in the respective parallel lines (in $\reals^3$)
\begin{align*}
q\cdot v_1 & = c_1 \\
q\cdot v_2 & = c_2,\quad\text{and} \\
p\cdot v_1 & = u_1 \\
p\cdot v_2 & = u_2 .
\end{align*}
As in the planar case, these lines need not be distinct, so $s = \nu^2$, where $\nu=\nu(S)$ is the
maximum size of a collinear subset of $S$.

For estimating $q$, a simplified variant of the analysis just given shows that the maximum number of lines
$h_{p,q}$ that lie in a common hyperplane is $\mu^2$, where $\mu=\mu(S)$ is the maximum number 
of points of $S$ that lie in a common plane (in $\reals^3$). The situation is more involved for quadrics.
Let $Q$ be a quadric, whose equation is given by $(\xi,t,1)A(\xi,t,1)^T = 0$, for a suitable
$5\times 5$ symmetric matrix $A$. Then $h_{p,q}$ is contained in $Q$ if
$$
(q-tp,t,1)A(q-tp,t,1)^T = 0 
$$
for every $t$. That is, we must have
\begin{align*}
(q,0,1)A(q,0,1)^T & = 0 \\
(p,-1,0)A(p,-1,0)^T & = 0 \\
(p,-1,0)A(q,0,1)^T & = 0 .
\end{align*}
That is, $p$ lies on a quadric $Q_0$ in 3-space, $q$ lies on another ``similar'' quadric 
(that has the same quadratic part as $Q_0$), and $p$ and $q$ satisfy a bilinear equality induced by $Q_0$
(the third equation given above).
We can therefore bound, pessimistically, the number of lines $h_{p,q}$ that lie on a quadric by $\kappa^2$,
where $\kappa = \kappa(S)$ is the maximum number of points of $S$ that lie in a common quadric in 3-space
(it looks like the actual bound should be smaller).
That is, we have $q \le \max\{\mu^2,\kappa^2\}$.

Substituting the bounds on $s$ and $q$ in (\ref{plinc4}), we get
$$
I(H,L) = O\left( 2^{c\sqrt{\log m}} (m^{2/5}N^{4/5} + m) + m^{1/2}N^{1/2}(\mu^{1/2}+\kappa^{1/2}) + m^{2/3}N^{1/3}\nu^{2/3} + N \right) .
$$
Arguing as above, with $N=n^2$, this implies that the number $M_{\ge t}$ of $t$-rich homotheties satisfies
\begin{equation} \label{gek3}
M_{\ge t} = O\left( \frac{2^{O(\sqrt{\log n})}n^{8/3}}{t^{5/3}} + \frac{n^2(\mu+\kappa)}{t^2} + \frac{n^2\nu^2}{t^3} + \frac{n^2}{t} \right) .
\end{equation} 
Note that the first (resp., second) term in (\ref{gek3}) is smaller than the planar counterpat term
$O(n^3/t^2)$ in (\ref{gek}) when $t\ll n$ (resp., when $\mu, \kappa \ll n$); the third and fourth terms 
in (\ref{gek3}) are the same as in (\ref{gek}). That is, when $t, \mu, \kappa \ll n$ we get a smaller bound
on the number of $t$-rich homotheties in 3-space than we get in the plane.

\paragraph{Lower bound for distinct homothety classes.}
Let $S$ be a set of $n$ points in $\reals^3$, and let $k\ge 3$ be an integer parameter.
As in the planar case, we estimate the number of distinct equivalence classes of $k$-element 
subsets of $S$ under homotheties, via the set $Q$ of all pairs of equivalent $k$-tuples.
We have the same upper lower bound $|Q| = \Omega\left( n^{2k}/x \right)$, where $x$
is the number of equivalence classes. For an upper bound on $|Q|$, we note, as before,
that every homothety that maps exactly $t$ points of $S$ to $t$ other points generates 
at most $t^k$ elements of $Q$, implying that
$$
|Q| \le \sum_{t\ge k} t^k M_t = O\left( k^k M_{\ge k} + \sum_{t \ge k+1} t^{k-1} M_{\ge t} \right) ,
$$
where $M_t$ (resp., $M_{\ge t}$) is the number of homotheties that map exactly $t$ 
(resp., at least $t$) points of $S$ to other points of $S$.
We could have used the upper bound (\ref{gek3}) on $M_{\ge t}$ to estimate this expression, 
but, as can be easily verified, and as we have already forewarned, we do not get any improvement over the planar case.
In fact, the bound is slightly worse because of the presence of the factor $2^{O(\sqrt{\log n})}$ in (\ref{gek3}).

We can get the same lower bound as in the plane by first arguing that the upper bound in (\ref{gek})
also holds in three (and in fact in any higher) dimensions. This is because a generic projection of the 
$t$-rich homotheties and of the lines $h_{p,q}$ onto some generic 3-space has the properties that 
(i) incidences are preserved, (ii) no pair of lines and no pair of points (homotheties) have coinciding images, 
and (iii) no plane contains more than $\nu^2$ lines. That genericity implies the first two properties is clear, 
and that it implies property (iii) requires a short and easy argument that we omit here.

Hence, using this reduction, we obtain that the lower bounds in Theorem~\ref{dist:hom} hold in any
dimension $d\ge 3$ too. They are worst-case tight for $k\ge 4$, and are tight 
for $k=3$ when $\nu = O(n/\sqrt{\log n})$. 

\paragraph{Similarities.}
Going back to the plane, we
remark that the case of similarities is also amenable to the technique of Elekes, which, in this case
reduces the problem to incidence questions points and lines in the complex plane. Specifically,
if we regard the real plane as the complex line $\cplx$, a similarity tranformation in the plane is
a linear transformation $z\mapsto \xi z+\eta$, for $\xi, \eta\in\cplx$, and vice versa. Indeed, multiplying by $\xi$ 
represents rotation and scaling about the origin, and $\eta$ is the subsequent translation vector.
We thus represent similarities as points in $\cplx^2$.
The locus $h_{p,q}$ of all similarities that map $p$ to $q$ is the complex line $p\xi + \eta = q$
in the $\xi\eta$-plane. 

Using the extension of the Szemer\'ei-Trotter incidence bound to the complex plane,
due to T\'oth~\cite{To} and to Zahl~\cite{Za}, one can show, in a completely straightforward manner,
that, for a set $S$ of $n$ points in the plane, the number of $k$-rich similarities (those that
map at least $k$ points of $S$ to other points of $S$) is $O(n^4/k^3)$, a bound already derived by
Solymosi and Tardos~\cite{SoTa}, using a different (more ``elementary'') technique, and also noted by Rudnev~\cite{Rud}.
This is turn implies that the number of pairwise non-similar triangles determined by $n$ points 
in the plane is $\Omega(n^2/\log n)$; again, see \cite{Rud,SoTa}. Exactly the same machinery
can be used to derive lower bounds on the number of pairwise non-similar $k$-tuples determined 
by $n$ points in the plane. Although they do not state it explicitly, the papers just cited 
could have also obtained this extension with their technique.

\paragraph{A final note.}
The geometric and algebraic structure of homotheties is much simpler than that of rigid motions 
(and of similarities). It is therefore somewhat surprising that the new bounds derived in this 
note have not been obtained earlier, by a more ``direct'' geometric approach, such as in \cite{SoTa}. 
While the application of the algebraic machinery to homotheties, as presented in this paper, is 
interesting and pleasing (to us), we honestly have no idea whether it is indeed necessary, and 
leave it as an interesting open problem to come up with more ``elementary'' proofs.

\end{document}